\documentclass[aps,
prl,showpacs,showkeys,
twocolumn,
floatfix,
superscriptaddress
]{revtex4}

\usepackage{graphicx}
\usepackage{multirow}
\usepackage{longtable}
\usepackage{rotating}
\usepackage{lipsum}

\usepackage{amsmath,amstext,amsfonts,amsbsy,amssymb,amscd,bbm,epsfig}

\newcommand{\be}{\begin{equation}}
\newcommand{\ee}{\end{equation}}
\newcommand{\bea}{\begin{eqnarray}}
\newcommand{\eea}{\end{eqnarray}}

\allowdisplaybreaks[4] 

\begin{document}

\begin{titlepage}
\renewcommand{\thefootnote}{\alph{footnote}}

\title{Impact of nuclear effects on the
  extraction of neutrino oscillation parameters}

\author{P.~Coloma}

\author{P.~Huber}
\affiliation{Center for Neutrino Physics, Virginia Tech, Blacksburg, VA 24061, USA}
\date{\today}

\begin{abstract}
We study the possible impact of nuclear effects and final state
interactions on the determination of the oscillation parameters due to
mis-reconstruction of non-quasi-elastic events as quasi-elastic events
at low energies. We analyze a $\nu_\mu$ disappearance experiment using
a water \v{C}erenkov detector. We find
that, if completely ignored in the fit, nuclear effects can induce a
significant bias in the determination of atmospheric oscillation
parameters, particularly for the atmospheric mixing
angle. Even after inclusion of a near detector a bias in the
determination of the atmospheric mixing angle comparable to the
statistical error remains.
\end{abstract}

\maketitle

\end{titlepage}

Neutrino oscillation is firm evidence for physics beyond the Standard
Model and therefore, a rich program of neutrino oscillation
experiments is ongoing and more ambitious projects are planned for the
future. The eventual goal is to obtain sufficient precision as to be
able to uncover the mechanism responsible for neutrino masses and
mixing. In a nut shell, neutrino physics is evolving to become a
precision science.  In order to reach that goal, neutrino-nucleus
interaction cross sections have to be known with sufficient
accuracy. So far only very few studies exist which establish a
quantitative connection between uncertainties on neutrino cross
sections and the resulting induced error in the determination of
neutrino mixing parameters, see {\it e.g.}  Refs.~\cite{Huber:2007em,Coloma:2012ji,Itow:2002rk,Harris:2004iq}.

Recent experimental results on neutrino cross sections however, for
instance from MiniBooNE~\cite{AguilarArevalo:2010zc} or
MINER$\nu$A~\cite{Fiorentini:2013ezn}, indicate that not only the
total cross sections have large uncertainties but also the energy
dependence and energy distribution of secondary particles is not well
understood. The reason presumably lies in neglected nuclear effects
and/or final state interactions -- the fact that nucleons are bound
inside the nucleus has multiple implications: (1) the initial and
final state densities are modified; (2) many-particle correlations
play an increased role; and (3) any reaction product has to make it
out of the nucleus in order to be observed in the detector. For
brevity we will refer to these phenomena collectively as nuclear
effects. Obviously, a closed form description of this system is beyond
our current abilities. Many approximate calculations exist, but only
very few have been tested rigorously against data.  A number of
studies have addressed the impact of nuclear effects on oscillation
analyses.  In Refs.~\cite{FernandezMartinez:2010dm, Meloni:2012fq} the
expected sensitivity of some oscillation experiments were presented
for different assumptions of the nuclear model. No final state
interactions were considered, though, and the nuclear model was
assumed to be known by the time the data are analyzed. A different
problem was considered in
Refs.~\cite{Lalakulich:2012hs,Martini:2012fa,Nieves:2012yz}, where a
\emph{qualitative} description of the impact that final state
interactions and multinucleon interactions may have on the event distribution was presented. In this
work, we attempt to provide a \emph{quantitative}
estimate of the bias that the uncertainties on nuclear effects may
induce in the determination of neutrino oscillation parameters, following a similar approach as in Refs.~\cite{Lalakulich:2012hs,Martini:2012fa,Nieves:2012yz}.
 
\begin{figure}[t!]
\begin{center}
  \includegraphics[width=\columnwidth]{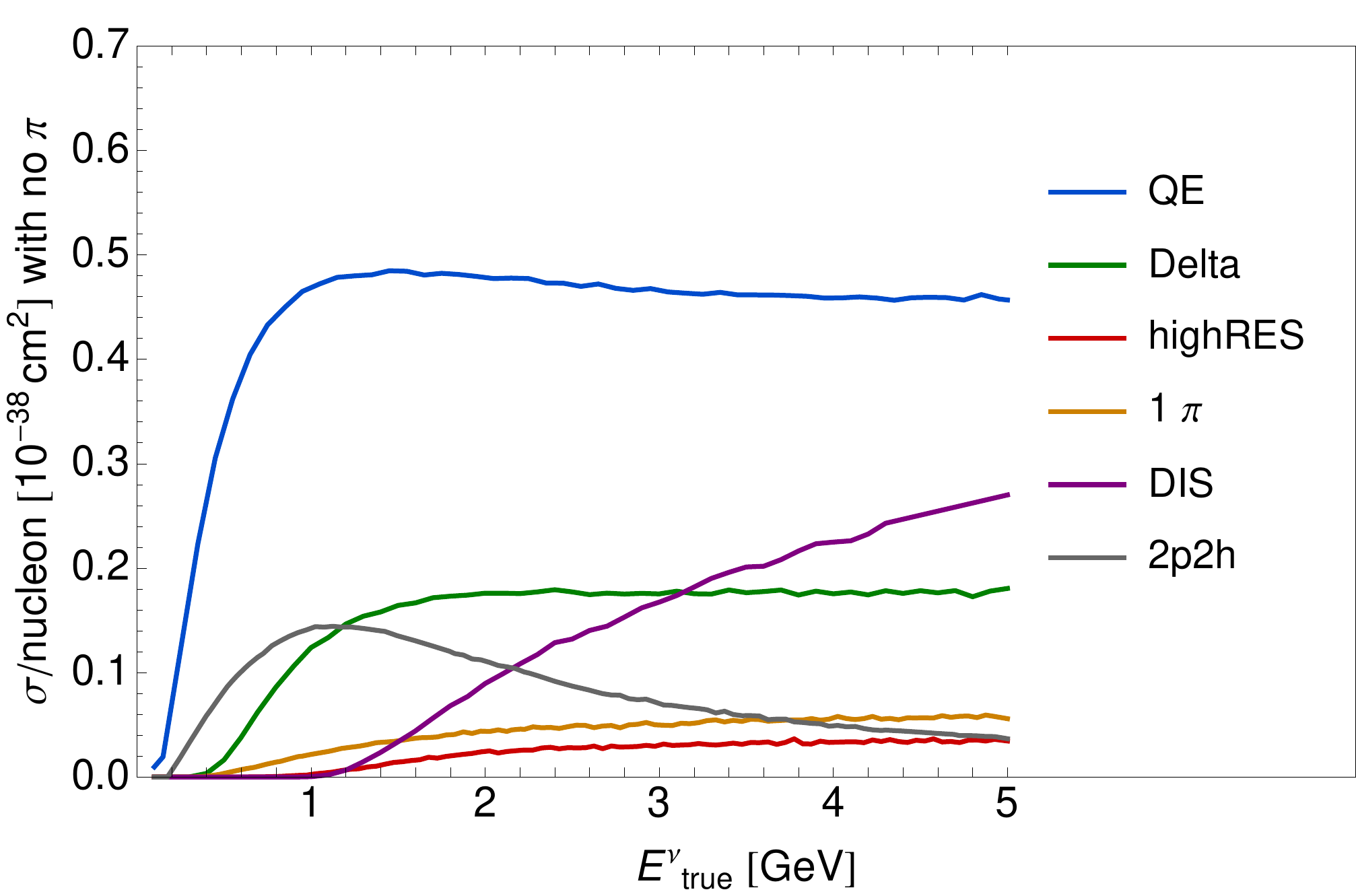}
\end{center}
  \caption{ Neutrino interaction cross section per nucleon for several
    processes in $^{16}$O with no pions in the final state, as a
    function of the true neutrino energy. The labels in the legend
    indicate: quasi-elastic (``QE''); $\Delta$ production (``Delta'');
    one pion production (``$1\pi$''); production of higher resonances
    (``highRES''); deep inelastic scattering (``DIS'') and
    two-particle-two-hole interactions (``2p2h''). }
\label{fig:xsec}
\end{figure}

We focus the analysis on the so-called atmospheric oscillation
parameters, $\theta_{23}$ and $\Delta m^2_{31}$ using quasi-elastic
charged current (CC-QE) $\nu_\mu$ events from a $\nu_\mu$ beam, a
so-called disappearance experiment.  Generally speaking, in a neutrino
oscillation experiment the amplitude of the oscillation provides a
measurement of the mixing angle, while the energy dependence provides
a measurement of the squared mass splitting. From this statement it is
obvious that a correct identification of the neutrino energy is
crucial to determine the mass splitting. However, a wrong
identification of the neutrino energy can result in a pile-up of the
events at different neutrino energies, which would translate into a
wrong determination of the mixing angle as well. For a CC-QE event,
where the charged lepton produced in the final state is most readily
observed, the neutrino energy is usually reconstructed using the
kinematic variables of the charged lepton only.  In the absence of
nuclear effects, the number of events with neutrino energy $E_i$ that
are truly QE can be easily computed as:
\begin{equation}
N^{QE}_i= \sigma^{QE}(E_i)\phi(E_i)P_{\mu\mu}(E_i) \, ,
\label{eq:noMM}
\end{equation}
where $P_{\mu\mu}$ is the $\nu_\mu$ disappearance oscillation
probability, $\phi$ is the flux and $\sigma$ is the cross section for
QE events, which is shown by the blue line in Fig.~\ref{fig:xsec}.

Let us consider now the case of a CC neutrino interaction that is not
QE. Usually, these interactions are discarded from the event sample if
another charged particle (for example, a pion) is observed in the
final state. However, there is a certain probability that the produced
pion is absorbed by the nucleus and is therefore not detected. In this
case, the only observable particle in the final state will be the
charged lepton and consequently this event will be added to the QE
sample. In addition, since the event was not purely QE and a particle
in the final state was missed, this will most likely lead to a
reconstructed energy smaller than the true incident neutrino
energy. As a consequence, each bin in reconstructed neutrino energy
will receive contributions from events that took place at different
true neutrino energies:
\begin{widetext}
\begin{eqnarray}
N^{QE-like}_i & = & \sum_j M^{QE}_{ij} N^{QE}_j + \sum_{non-QE}\sum_j
M^{non-QE}_{ij} N^{non-QE}_j \label{eq:MM} \\ & = & \sum_j M^{QE}_{ij}
\sigma^{QE}(E_j)\phi(E_j)P_{\mu\mu}(E_j) + \sum_{non-QE} \sum_j
M^{non-QE}_{ij} \sigma^{non-QE}_{0\pi}(E_j)\phi(E_j)P_{\mu\mu}(E_j) \,
. \nonumber
\end{eqnarray}
\end{widetext}
Here, the matrices $M_{ij}$ account for the probability that an event
with a true neutrino energy in the bin $j$ ends up being reconstructed
in the energy bin $i$. The matrix $M_{ij}^{QE}$ is mostly diagonal and
just adds a certain, quasi-Gau\ss ian, smearing over
Eq.~\ref{eq:noMM}. However, for non-QE this is not going to be the
case. Different migration matrices are obtained depending on the
particular interaction that has initially taken place. Therefore, a
sum is performed over the different non-QE processes that take place
in the detector with no second charged particle, {\it i.e.} pion, in
the final state. The neutrino interaction cross sections on $^{16}$O
with no pions in the final state, $\sigma_{0\pi}$, are shown in
Fig.~\ref{fig:xsec} for all the processes under consideration in this
work. Migration matrices for $^{16}$O have been produced for each of
these processes following Ref.~\cite{Lalakulich:2012hs}.  The
GiBUU~\cite{Buss:2011mx} transport model has been used to generate
both the migration matrices and the cross sections used in this work,
see Ref.~\cite{Buss:2011mx} for a useful review of transport models
and details about GiBUU.

\begin{figure}[t!]
\begin{center}
  \includegraphics[width=0.95\columnwidth]{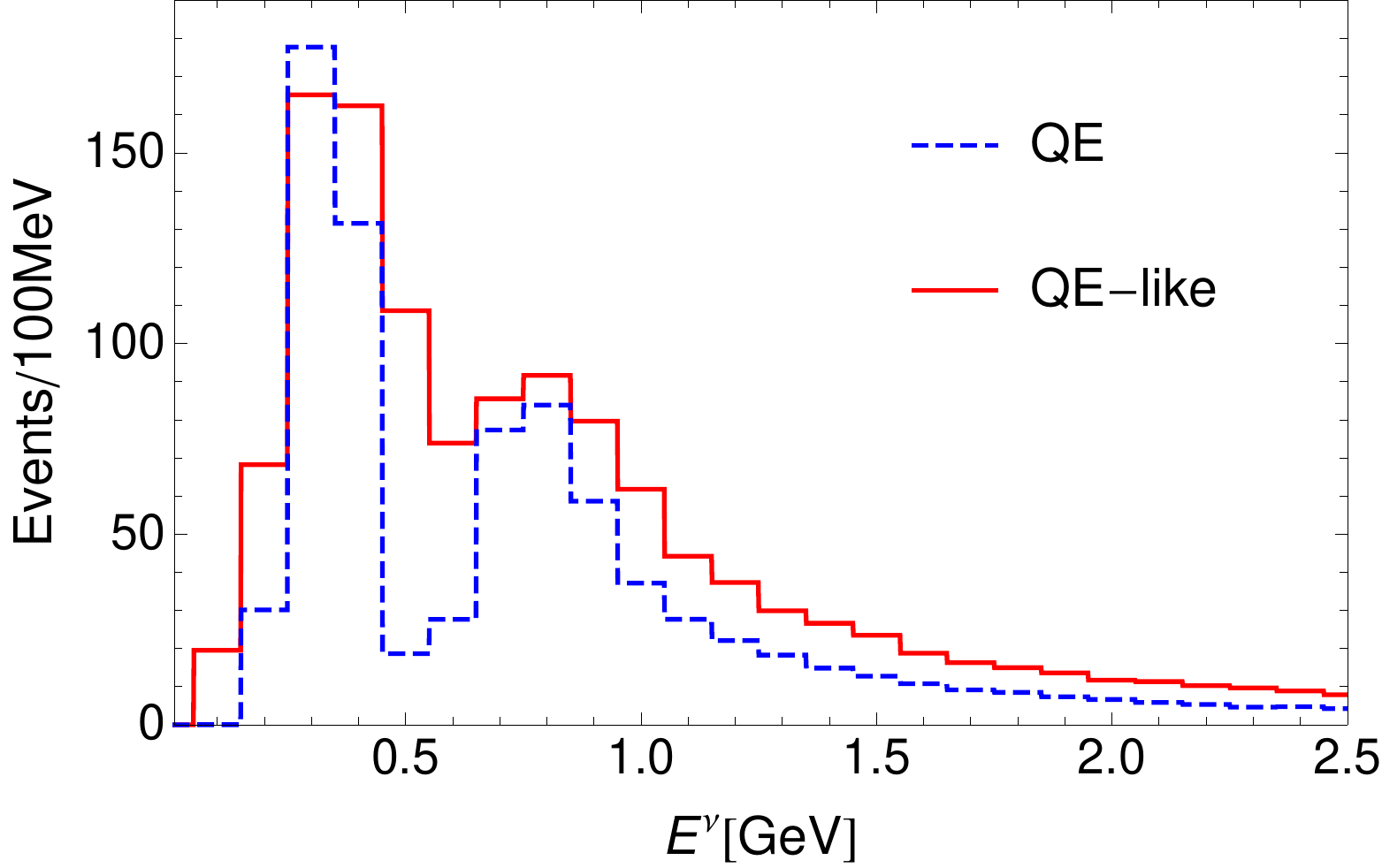} 
  \caption{ Distribution of QE-like and truly-QE events as a function
    of reconstructed neutrino energy. Signal efficiencies have already
    been accounted for. Background events are not included. }
\label{fig:events}
\end{center}
\end{figure}

 For the sake of simplicity we will use as input values for our
 analysis $\theta_{23}=45^\circ$ and $\Delta
 m^2_{31}=2.45\times10^{-3}\,\mathrm{eV}^2$ and chose the other
 oscillation parameters according to
 Ref.~\cite{GonzalezGarcia:2012sz}. To illustrate the potential issues
 arising from nuclear effects, we choose as example a low energy
 neutrino oscillation experiment, where a muon neutrino
 beam with a mean energy of $600$ MeV is aimed at a water \v{C}erenkov
 detector, mainly sensitive to QE events only. In particular, only
 events with one charged particle above \v{C}erenkov threshold are
 selected as signal, so-called single-ring events. The single-ring
 event criterion is very easy to implement in our calculation and the
 effect of non-QE events on the QE event sample therefore can be
 estimated without a detailed detector simulation. Many other
 experiments have detectors which also are sensitive to hadronic
 energy deposition in the detector and in some cases even will be able
 to reconstruct proton tracks (at least for a subset of events). For
 these detectors, a study of nuclear effects on energy reconstruction
 requires a detailed detector simulation, which is beyond the scope of
 this work. We consider only the $\nu_\mu\rightarrow \nu_\mu$
 disappearance channel. The neutrino flux is the same as in Ref.~\cite{T2K}. 
 For 5 years of data taking assuming a beam power of 750 kW, our calculation 
 yields an approximate number of $\sim850$ true QE
 events and about $\sim 1300$ QE-like events. The event distributions
 as a function of the neutrino energy can be seen in
 Fig.~\ref{fig:events} for both cases. The data is divided into 100
 MeV bins between 0.2 and 2 GeV. Energy dependent signal efficiencies for 
 a water \v{C}erenkov detector, following Ref.~\cite{Huber:2009cw}, have been included.

\begin{figure*}[t!]
\begin{center}
\begin{tabular}{cc}
  \includegraphics[width=\columnwidth]{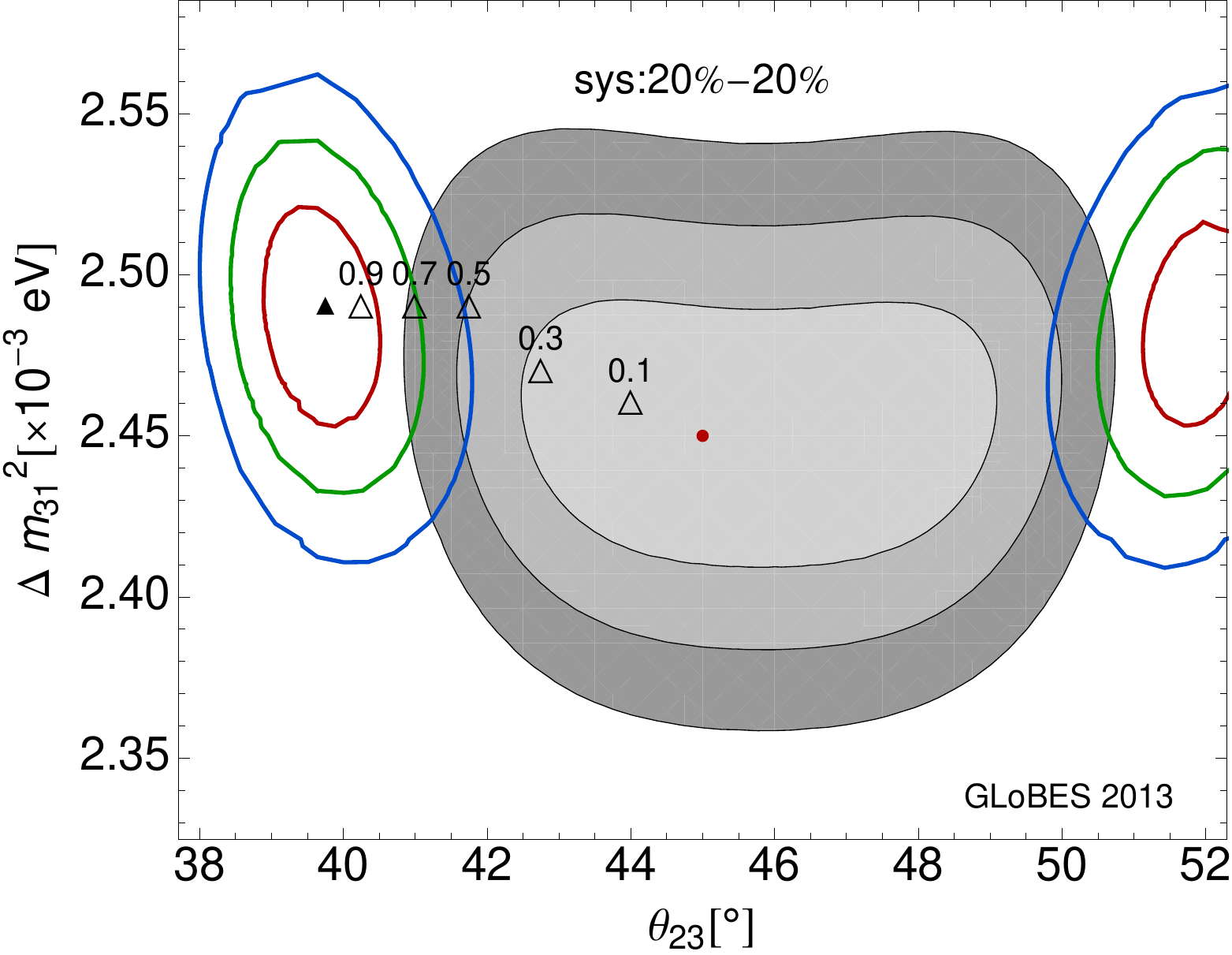} &
  \includegraphics[width=\columnwidth]{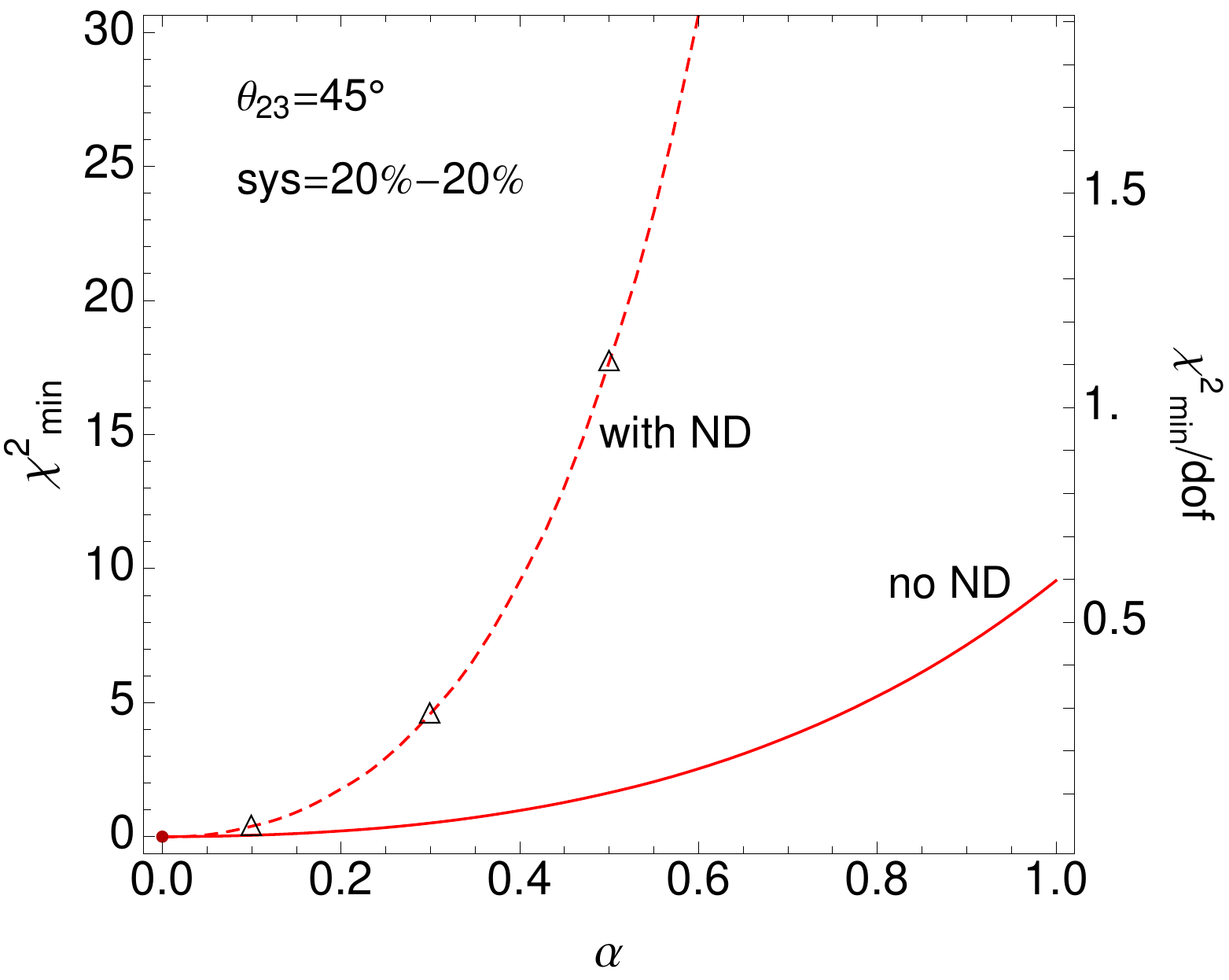}
\end{tabular}
\end{center}
  \caption{ Left: Confidence region in the $\theta-\Delta m^2$ plane
    for 2 d.o.f. for different scenarios. Gray shaded areas show the
    results assuming the nuclear model is perfectly known. The lines
    depict the 1, 2 and 3\,$\sigma$ regions for a fit taking $\alpha=1$,
    where $\alpha$ represents the amount of migration due to nuclear
    effects that is being neglected in the fit, see text for
    details. A near detector is included in both cases.  The triangles
    indicate where the best fit lies for the region enclosed by the
    colored lines as $\alpha$ is increased from 0 (red dot,
    corresponding to the true input value) to 1 (filled black
    triangle). Right: Minimum $\chi^2$ as a function of $\alpha$. For each line, the minimum value of the $\chi^2$ is computed as the value of $\alpha$ is progressively increased from 0 to 1. The dashed (solid) line shows the result with(out) a near detector (ND). For illustration purposes, some of the triangles in the left-hand panel (which correspond to the results including a near detector) are explicitely shown in this panel as well.
  }
\label{fig:alpha}
\end{figure*}

The uncertainties of the other oscillation parameters are negligible
for the purpose of this work, and therefore we keep them fixed in the
fit. Systematic uncertainties, on the other hand, are relevant and two
types of systematic errors are included in the analysis: a $20\%$
normalization error, bin-to-bin \emph{correlated}; and a $20\%$ shape
error, bin-to-bin \emph{uncorrelated}. A binned poissonian $\chi^2_{i,D}$ is computed taking the signal rates per energy bin $i$ and detector $D$ as: $S_{i,D}(\theta,\xi)= (1+\xi_n + \xi_{\phi,i})N_{i,D}(\theta)$, where $\theta$ indicates the dependence on the oscillation parameters, and $ \xi_{\phi,i}$ and $\xi_n $ stand for the nuisance parameters associated to flux and normalization uncertainties, respectively.
The final $\chi^2$ reads:
\begin{equation}
\chi^2 = min_{\xi} \left\{ 
\sum_{D,i} \chi^2_{i,D}(\theta;\xi) + \left(\dfrac{\xi_{\phi,i}}{\sigma_\phi}\right)^2 + \left(\dfrac{\xi_{n}}{\sigma_{n}}\right)^2 
\right\} \, ,
 \nonumber
\end{equation}
where the first term corresponds to the binned poissonian $\chi^2$, and $\sigma_k$ indicate the prior systematic uncertainties assumed (20\% in all cases). More details on the $\chi^2$ implementation can be found in Ref.~\cite{Coloma:2012ji}. We simulate an ideal near detector
placed sufficiently far away from the source so that the observed
spectrum is the same as in the far detector, {\it i.e.} we assume an
energy-independent near/far ratio. Identical signal and background
efficiencies are used as well. It should be stressed that, in real
world cases these conditions are likely not satisfied and the
usefulness of the near detector is expected to be reduced.  A
Poissonian $\chi^2$, where systematic errors are fully correlated
between near and far detectors, is considered. We use a modified
version of the GLoBES software~\cite{Huber:2004ka,Huber:2007ji}, see
Ref.~\cite{Coloma:2012ji} for details. In the fit, the \emph{true}
distribution of events is always computed according to
Eq.~\ref{eq:MM}. Two possible extreme situations arise:
\begin{enumerate}
\item nuclear effects are \emph{completely ignored}, and we try to fit
  the true rates with the expected event rates computed from
  Eq.~\ref{eq:noMM};
\item nuclear effects are \emph{perfectly known}, and the fit is done
  computing the expected event rates using Eq.~\ref{eq:MM}.
\end{enumerate}
This is shown in the left panel in Fig.~\ref{fig:alpha}. As it can be
seen from the fit, the very different distributions in the number of
events in Fig.~\ref{fig:events} lead to a significant shift in the
best fit for the oscillation parameters, shown by the black triangle.

In reality, one is likely to be in-between these two extremes of no
versus perfect knowledge of nuclear effects. It is very difficult to
quantify the ``error'' on models of nuclear effects, since they are
generally not the result of a well-controlled expansion in some small
parameter. One possible way to address this question from a
phenomenological point of view is to introduce a parameterization
which allows to connect the two extremes in a continuous fashion
\begin{equation}
N^{\textrm{test}}_i (\alpha) = \alpha \times N^{QE}_i + (1-\alpha)\times N^{QE-like}_i  , \label{eq:alpha} 
\end{equation}
where $\alpha$ parameterizes the fraction of migration that is
neglected in the fit: $\alpha=0$ corresponds to Eq.~\ref{eq:MM} while
$\alpha=1$ corresponds to Eq.~\ref{eq:noMM}.  The inclusion of this 
parameter may be regarded as an additional systematic uncertainty. 
Similar uncertainties were included, for instance, in Ref.~\cite{Abe:2013xua}. 
However, it should be noted that in our fit it is not treated as a systematic 
uncertainty, since no marginalization over $\alpha$ is performed. Another possibility
would be to compute migration matrices according to Eq.~\ref{eq:MM}
using different nuclear models and take the spread of results as
measure of the error. For the following discussion $\alpha$ serves as
a proxy for the error of the nuclear model. The result of varying
$\alpha$ from 0 to 1 is shown in Fig.~\ref{fig:alpha}. The position of
the best fit for different values of $\alpha$ is shown in the left
panel by the empty triangles. As it can be seen from the figure, the
deviation of the best fit from the true input value is progressively
increased with the value of $\alpha$. In the right hand panel of
Fig.~\ref{fig:alpha} the increase of the minimum $\chi^2$ as function
of $\alpha$ is shown: clearly, an ``error'' in the nuclear model would
make the minimum $\chi^2$ get worse. A sufficiently large value of the
minimum of the $\chi^2$ (with respect to the effective number of
degrees of freedom) could eventually force to reject the fit. Note,
that rejecting the fit at the end of the experiment would still
indicate its failure. The right hand panel also indicates the effect
of a near detector: the solid line shows the result without a near
detector, whereas the dashed line shows the one with a near
detector. Generally, the near detector adds more tension to the fit if
the nuclear model is wrong and thus, serves as an indicator that
something is wrong. However, even for relatively large values of
$\alpha \sim 0.3-0.4$ the minimum value of $\chi^2$ would still be low
enough so that the fit may be accepted, even if a near detector is
included in the analysis. A value $\alpha=0.3$ still corresponds,
according to the left hand panel, to a $1\,\sigma$ bias in the
determination of the mixing angle. Therefore, it stands to reason that
adding a near detector may not be sufficient to completely cure the
problem: a successful experiment requires an accurate nuclear model,
where the accuracy of the model has been independently verified.

Our results indicate that, for an experiment observing only QE-like events, a $1\,\sigma$ bias
in the determination of $\theta_{23}$ could result from errors on the
nuclear model even when taking full advantage of the near
detector. This is a first study on the quantitative impact of nuclear
effects on the determination of oscillation parameters. This type of
study should be extended to experiments which can observe hadronic
activity in the detector, as well as to appearance experiments, in
particular in the context of leptonic CP violation measurements.

\acknowledgements

 We are particularly indebted to O.~Lalakulich and U.~Mosel for many
 useful discussions relating to GiBUU. We would also like to thank
 C.~Mariani, K.~McFarland and J.~Morfin for useful discussions. This
 work has been supported by the U.S. Department of Energy under award
 number \protect{DE-SC0003915}.


%
\end{document}